\documentclass[epj]{webofc}
\usepackage[varg]{txfonts}   % Web of Conferences font
\usepackage{hyperref}

\renewcommand*{\eqref}[1]{Eq.~(\ref{eq:#1})}

\newcommand*{\figref}[1]{Fig.~(\ref{fig:#1})}
\newcommand*{\figlab}[1]{\label{fig:#1}}

\wocname{\includegraphics[width=0.25cm,clip]{logoARENA} ARENA2018}
\wocname{ARENA2018}
\woctitle{ARENA2018}
\woctitle{\includegraphics[width=0.25cm,clip]{logoARENA} ARENA2018}
\begin{document}
\title{Present status and prospects of the Tunka Radio Extension}

\author{%\firstname{Galileo} \lastname{Galilei}\inst{1,3}\fnsep\thanks{\email{revues@edpsciences.org}}
%\and \firstname{Isaac} \lastname{Newton}\inst{2}
%\and \firstname{Albert} \lastname{Einsteini}\inst{3}
\firstname{D.}~\lastname{Kostunin}\inst{1},
\firstname{P.A.}~\lastname{Bezyazeekov}\inst{2},
\firstname{N.M.}~\lastname{Budnev}\inst{2},
\firstname{D.}~\lastname{Chernykh}\inst{2},
\firstname{O.}~\lastname{Fedorov}\inst{2},
\firstname{O.A.}~\lastname{Gress}\inst{2},
\firstname{A.}~\lastname{Haungs}\inst{1},
\firstname{R.}~\lastname{Hiller}\inst{1}\fnsep\thanks{now at the University of Zürich},
\firstname{T.}~\lastname{Huege}\inst{1}\fnsep\thanks{also at Vrije Universiteit Brussel, Brussels, Belgium},
\firstname{Y.}~\lastname{Kazarina}\inst{2},
\firstname{M.}~\lastname{Kleifges}\inst{3},
\firstname{E.E.}~\lastname{Korosteleva}\inst{4},
\firstname{L.A.}~\lastname{Kuzmichev}\inst{4},
\firstname{V.}~\lastname{Lenok}\inst{1},
\firstname{N.}~\lastname{Lubsandorzhiev}\inst{4},
\firstname{T.}~\lastname{Marshalkina}\inst{2},
\firstname{R.}~\lastname{Monkhoev}\inst{2},
\firstname{E.}~\lastname{Osipova}\inst{4},
\firstname{A.}~\lastname{Pakhorukov}\inst{2},
\firstname{L.}~\lastname{Pankov}\inst{2},
\firstname{V.V.}~\lastname{Prosin}\inst{4},
\firstname{F.G.}~\lastname{Schröder}\inst{1,5},
\firstname{D.}~\lastname{Shipilov}\inst{2}
\and
\firstname{A.}~\lastname{Zagorodnikov}\inst{2}~
(Tunka-Rex Collaboration)
}

%\author{\firstname{First author} \lastname{First author}\inst{1,3}\fnsep\thanks{\email{Mail address for first author}} \and
%        \firstname{Second author} \lastname{Second author}\inst{2}\fnsep\thanks{\email{Mail address for second author if necessary}} \and
%        \firstname{Third author} \lastname{Third author}\inst{3}\fnsep\thanks{\email{Mail address for last author if necessary}}
%        % etc.
%}

\institute{
Institut für Kernphysik, Karlsruhe Institute of Technology (KIT), Karlsruhe, 76021 Germany \and
Institute of Applied Physics ISU, Irkutsk, 664020 Russia \and
Institut für Prozessdatenverarbeitung und Elektronik, Karlsruhe Institute of Technology (KIT) \and
Skobeltsyn Institute of Nuclear Physics MSU, Moscow, 119991 Russia \and
Department of Physics and Astronomy, University of Delaware, Newark, DE, USA
}

\abstract{%
The Tunka Radio Extension (Tunka-Rex) is a digital radio array operating in the frequency band of 30-80 MHz and detecting radio emission from air-showers produced by cosmic rays with energies above 100 PeV. 
The experiment is installed at the site of the TAIGA (Tunka Advanced Instrument for cosmic rays and Gamma Astronomy) observatory and performs joint measurements with the co-located particle and air-Cherenkov detectors in passive mode receiving a trigger from the latter. 
Tunka-Rex collects data since 2012, and during the last five years went through several upgrades. As a result the density of the antenna field was increased by three times since its commission. 
In this contribution we present the latest results of Tunka-Rex experiment, particularly an updated analysis and efficiency study, which have been applied to the measurement of the mean shower maximum as a function of energy for cosmic rays of energies up to EeV. 
The future plans are also discussed: investigations towards an energy spectrum of cosmic rays with Tunka-Rex and their mass composition using a combination of Tunka-Rex data with muon measurements by the particle detector Tunka-Grande.}
\maketitle
\section{Introduction}
Digital radio arrays are fast developing instruments for measuring ultra-high energy astrophysical messengers in PeV-EeV range~\cite{Schroder:2016hrv,Huege:2016veh}.
While the first and second generation setups were focused on technological developments and cosmic ray measurements, the next (third) generation radio array will aim on PeV-gamma~\cite{Schroeder_ARENA2018} EeV-neutrino~\cite{Martineau-Huynh:2015hae} detection.
In this work we review the status and highlight the latest results of the Tunka Radio Extension, part of the second generation setup mentioned above, which has proven the feasibility of ultra-high energy cosmic ray detection with a sparse radio arrays and obtained several important results, which have boosted the development of this technique.

\section{Tunka Radio Extension: history, location and equipment}

\begin{figure}[t]
\centering
\includegraphics[width=1.0\textwidth]{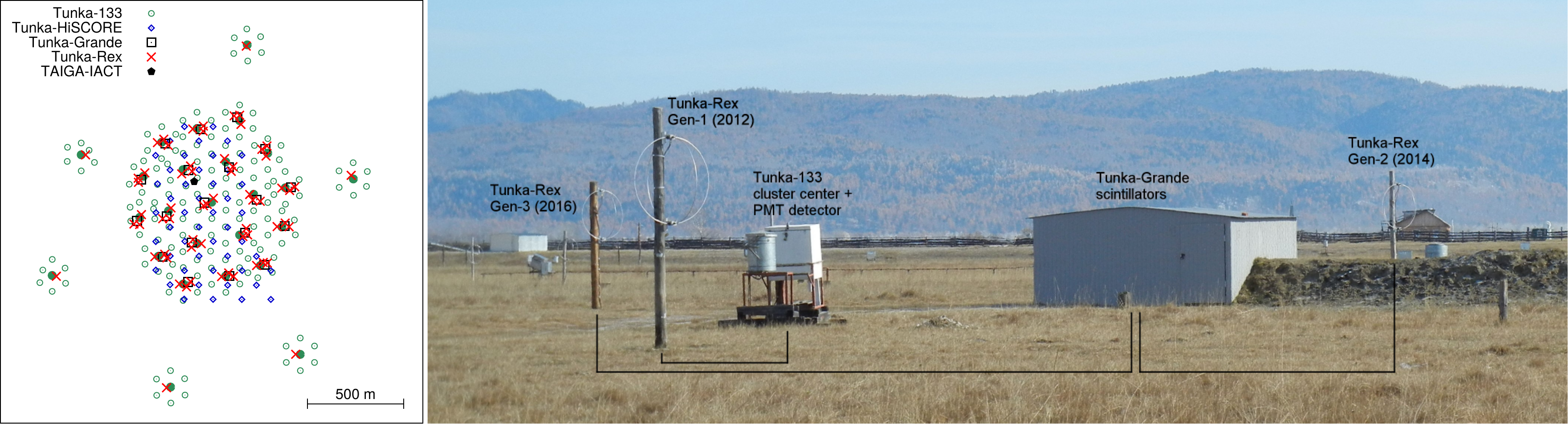}
\caption{\textit{Left:} layout of TAIGA facility. One can see, that cosmic-ray setups (Tunka-133, Tunka-Rex and Tunka-Grande) are grouped in clusters.
\textit{Right:} photo of single cosmic-ray cluster of TAIGA facility. Lines depict the cable connections between Tunka-Rex antennas and DAQ.}
\figlab{fig:cluster}
\end{figure}

\begin{figure}[t]
\centering
\includegraphics[width=1.0\textwidth]{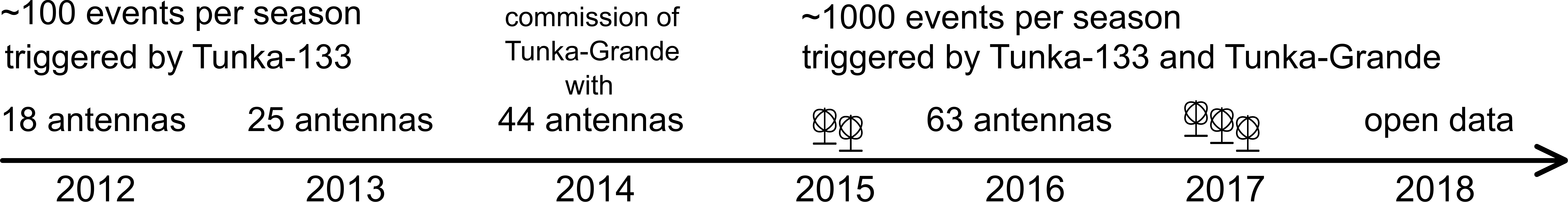}
\caption{Timeline of Tunka-Rex development.
The antenna array has been commissioned in 2012 with 18 antenna stations triggered by the Tunka 133 modules.
Since commission of Tunka-Grande in 2014-2015 Tunka-Rex recieves trigger from Tunka-Grande as well (during daytime measurements).
Starting from 2018, we are working on public access to Tunka-Rex software and data.}
\figlab{fig:timeline}
\end{figure}

Tunka Radio Extension (Tunka-Rex) is a digital antenna array located at the Tunka Advanced Instrument for cosmic rays and Gamma Astronomy (TAIGA) observatory~\cite{Budnev:2016btu}.
%~ The present status of TAIGA facility is shown in \figref{fig:cluster}.
TAIGA setups can be arbitrary divided in two main instruments: cosmic-ray instrument (Tunka-133~\cite{Prosin:2015voa}, Tunka-Rex~\cite{Bezyazeekov:2015rpa} and Tunka-Grande~\cite{Budnev:2015cha}) and gamma-ray instrument (Tunka-HiSCORE~\cite{Tluczykont:2017pin} and TAIGA-IACT~\cite{Yashin:2015lzw}).
In the left side of \figref{fig:cluster} one can see the layout of entire facility and note, that cosmic-ray setups are grouped in clusters: 19 clusters in dense core and 6 satellite clusters.
Each core cluster is equipped with 3 Tunka-Rex antenna stations, while satellite clusters contain single antenna stations and no Tunka-Grande scintillators.
In the right side of \figref{fig:cluster} one can see the photo of single cluster with corresponding detectors.

For the time being Tunka-Rex consists of 57 antenna stations located in the dense core of TAIGA (1~km\textsuperscript{2}) and 6 satellite antenna stations expanding area of array to 3~km\textsuperscript{2}.
Tunka-Rex has been commissioned in 2012 with 18 antenna stations triggered by air-Cherenkov array Tunka-133.
Each Tunka-Rex antenna station consists of two perpendicular active Short Aperiodic Loaded Loop Antennas (SALLA)~\cite{Abreu:2012pi} pre-amplified with Low Noise Amplifier (LNA).
Signals from antenna arcs are transmitted via 30~m coaxial cables to the analog filter-amplifier, which cuts frequency band to 30-80~MHz.
The selected signal is then digitized by the local data acquisition system (DAQ) with a 12 bit-sampling at a rate of 200 MHz; the data are collected in traces made of 1024 samples each.
Next years Tunka-Rex has been upgraded several times as well as TAIGA facility, which has been equipped with Tunka-Grande scintillator array providing trigger for Tunka-Rex since 2015.
One can see the timeline of Tunka-Rex development in \figref{fig:timeline}.

\section{Tunka-Rex calibration}
To reconstruct the electric field at the antenna it is necessary to know the hardware response of the antenna station, namely antenna pattern, and the gain and phase responses of the electronics.
The signal circuit of Tunka-Rex was calibrated in laboratory.
The antenna pattern and phase response were calculated with the simulation code NEC2~\cite{nec2}, then a calibration of the absolute gain was performed~\cite{TunkaRex_NIM_2015}.
The absolute amplitude calibration of the Tunka-Rex antenna station was performed with the same reference source as for LOPES~\cite{LOPES:2015eya}
which enabled us to perform a cross-check between KASCADE-Grande and Tunka-133 energy scales~\cite{Apel:2016gws}.

In Ref.~\cite{Kostunin:2017rbf} we suggested an approach for $X_\mathrm{max}$ reconstruction which uses the full information of the radio measurements, i.e. uses measured electric fields at the antennas (instead of only the maximum of signal amplitudes or signal powers).
Upon closer inspection, we have found that our phase calibration does not provide sufficient accuracy for exploiting this approach.
One can see the difference between simulated and reconstructed pulses in~\figref{fig:signal_diff}, which would introduce a significant systematic uncertainty in the analysis.
While phase response of Tunka-Rex chain is under investigation, the 10\% difference in simulated and measured gain can be evaluated by independent calibration against Galaxy performed on LOFAR~\cite{Nelles:2015gca}.
A preliminary study shows that, that bias between reference source VSQ 1000 and CoREAS simulations (\figref{fig:signal_diff}, 3) has the same behavior that between same source and calibration against Galaxy~\cite{Mulrey_ARENA2018}.

\begin{figure}[t]
\centering
\includegraphics[height=0.24\textwidth]{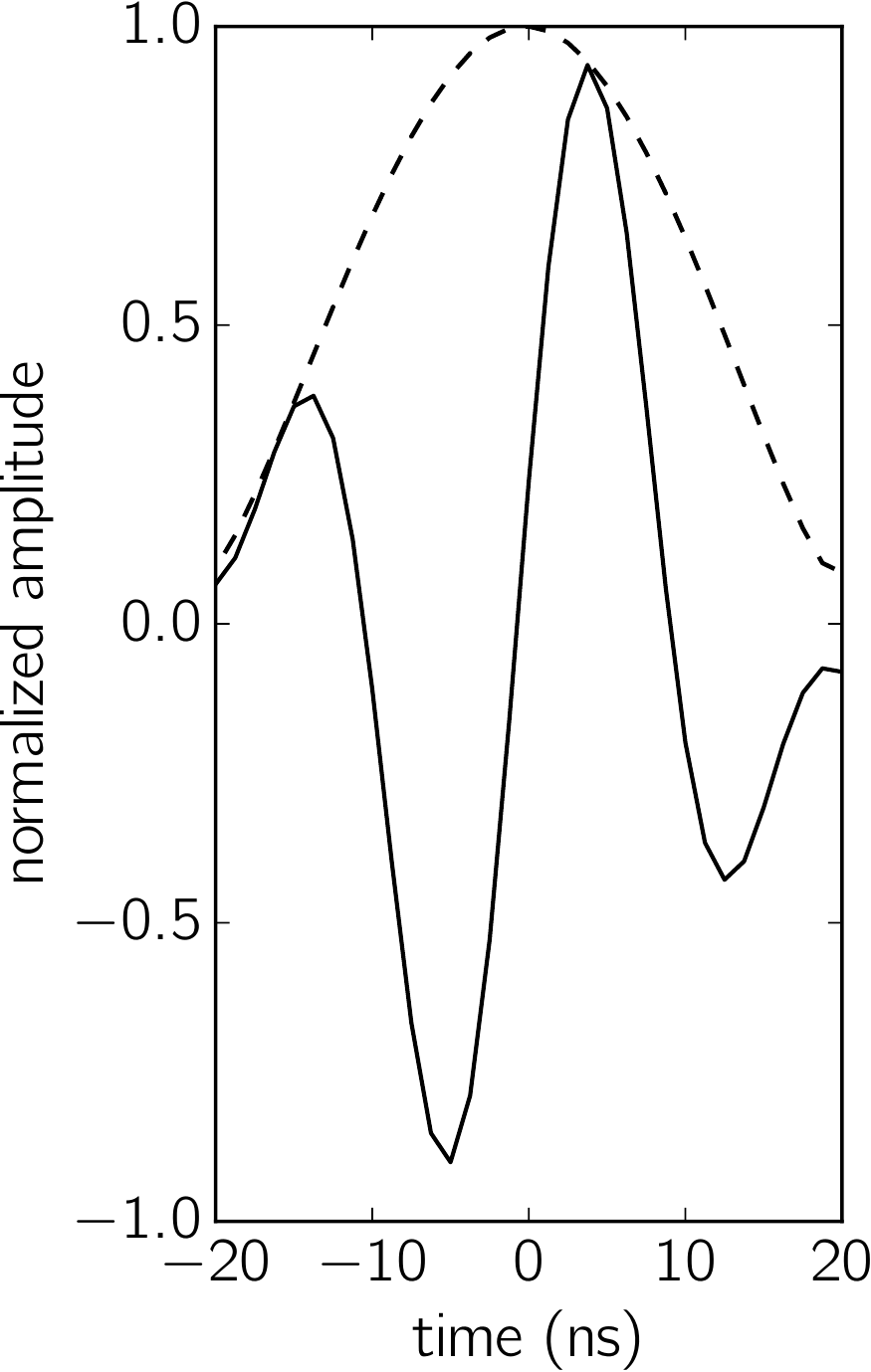}~~~
\includegraphics[height=0.24\textwidth]{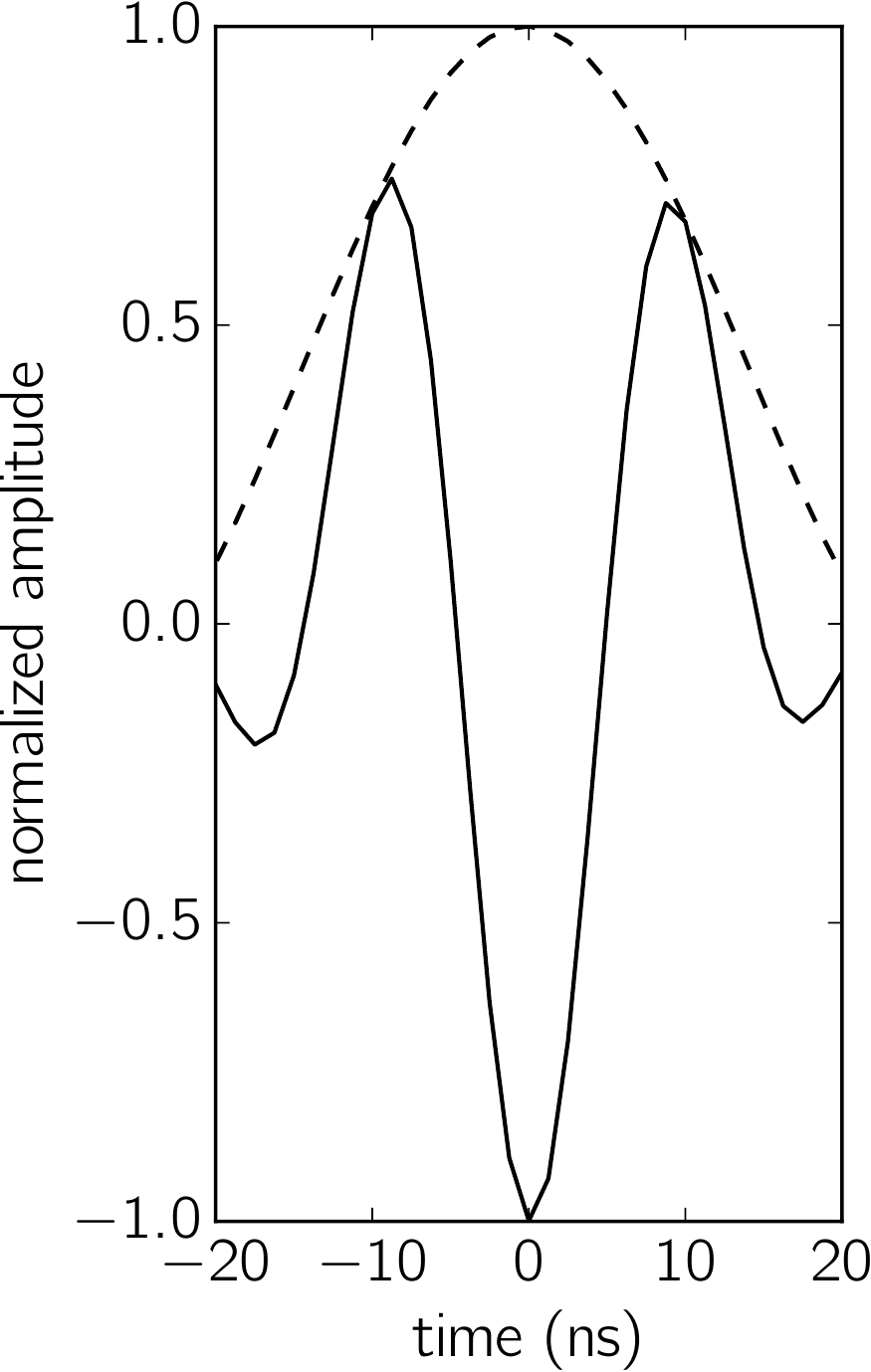}~~~
\includegraphics[height=0.24\textwidth]{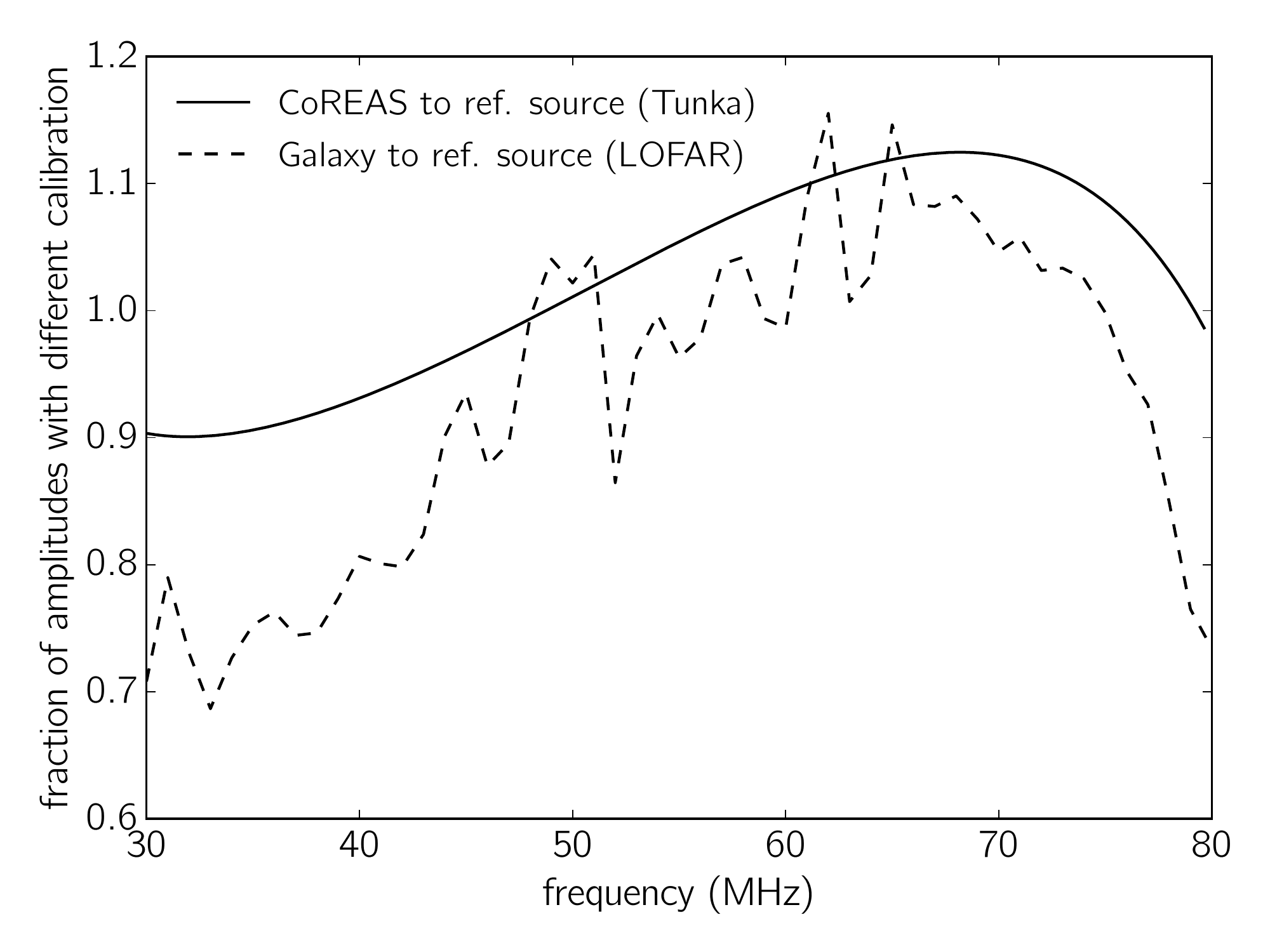}~~~
\includegraphics[height=0.24\textwidth]{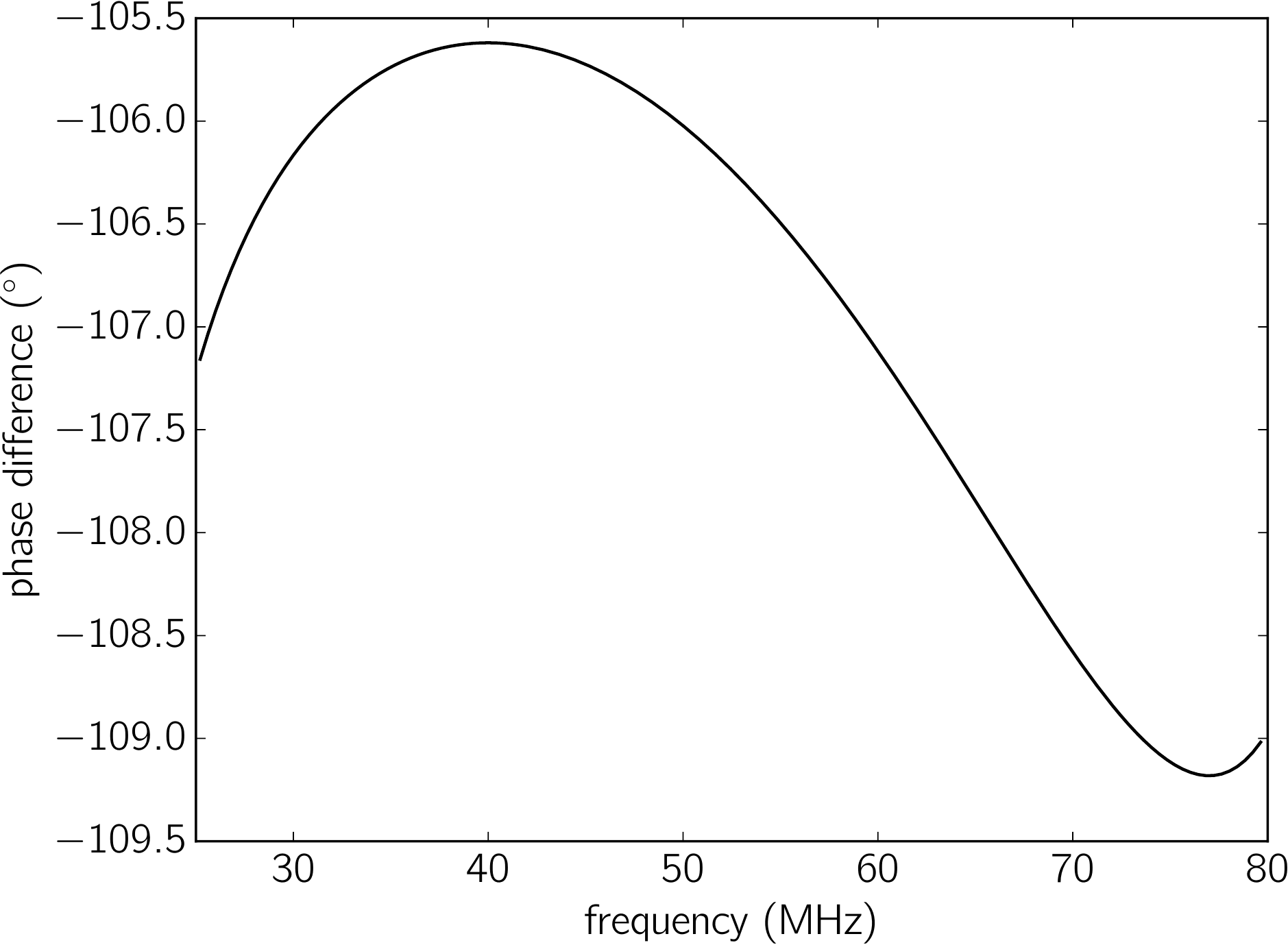}
\caption{From left to right: 
1) average of air-shower signals measured by Tunka-Rex; 
2) average of signals simulated with CoREAS and folded with Tunka-Rex hardware response; 
3) difference between measured and simulated spectral density of air-shower radio pulses compared to difference reconstructed by LOFAR using defferent calibrations~\cite{Mulrey_ARENA2018}; 
4) same for the phase response.
Dashed lines on (1,2) indicate envelope of the signal.}
\figlab{fig:signal_diff}
\end{figure}

\section{Updated signal processing and event reconstruction}

There are several improvements that have been introduced in Tunka-Rex signal analysis and event reconstruction in last years (a previous status of event reconstruction can be found in Ref.~\cite{Kostunin:2017bzd}), 
after upgrades at the TAIGA facility, when many additional RFI source appeared.
The most important improvements are described in the following items:
\begin{itemize}
\item \emph{The full width of a pulse} is limited to 50~ns.
Hereafter we define the pulse width as the distance between the two minima of the envelope closest to the peak (in~\figref{fig:signal_diff} the peak of the amplitude is at 0~ns, and the closest minima are at $-20$ and $20$~ns, i.e. the full width of the pulse is $40$~ns).
To prevent low-frequency RFI passing through the signal window all ``broad'' pulses (with a width of more than 50~ns) are omitted from the analysis.
The 50~ns window is determined from simulations, which showed that the width of air-shower signals is approximately 40-45~ns for the conditions of Tunka-Rex.  %(see Fig.~\ref{fig:signal_width}).
\item \emph{Sliding noise window}.
Experience has shown that the signal-to-noise ratio (SNR) estimation using a fixed noise window (slightly before the signal window, in case of Tunka-Rex) is affected by occasional RFI in the noise window.
To improve this estimation we use a sliding window of 500~ns and define the noise level as the smallest RMS in the entire trace within the noise window.
Since this value is systematically smaller than the average noise level, the threshold SNR was increased from $10.0$ to the value of ${\mathrm{SNR_{th}}=16.0}$.
\item \emph{Matched filtering and background suppression by neural networks}.
For the time being we are testing and implementing experimental techniques for lowering the threshold of the signal detection.
Since the pulse shape of air-shower radio signal can be described theoretically with CoREAS simulations (except phase bias, as mentioned above), one can use simulated pulses as templates for classical matched filtering.
Another approach is to train artificial neural network on background samples in order to create a library with filters, which will be used for the suppressing the background.
Both methods are tested on simulations and shown promising performance, the details can be read in Ref.~\cite{Shipilov_ARENA2018}.
\end{itemize}

\begin{figure}[t]
\centering
\includegraphics[height=0.35\textwidth]{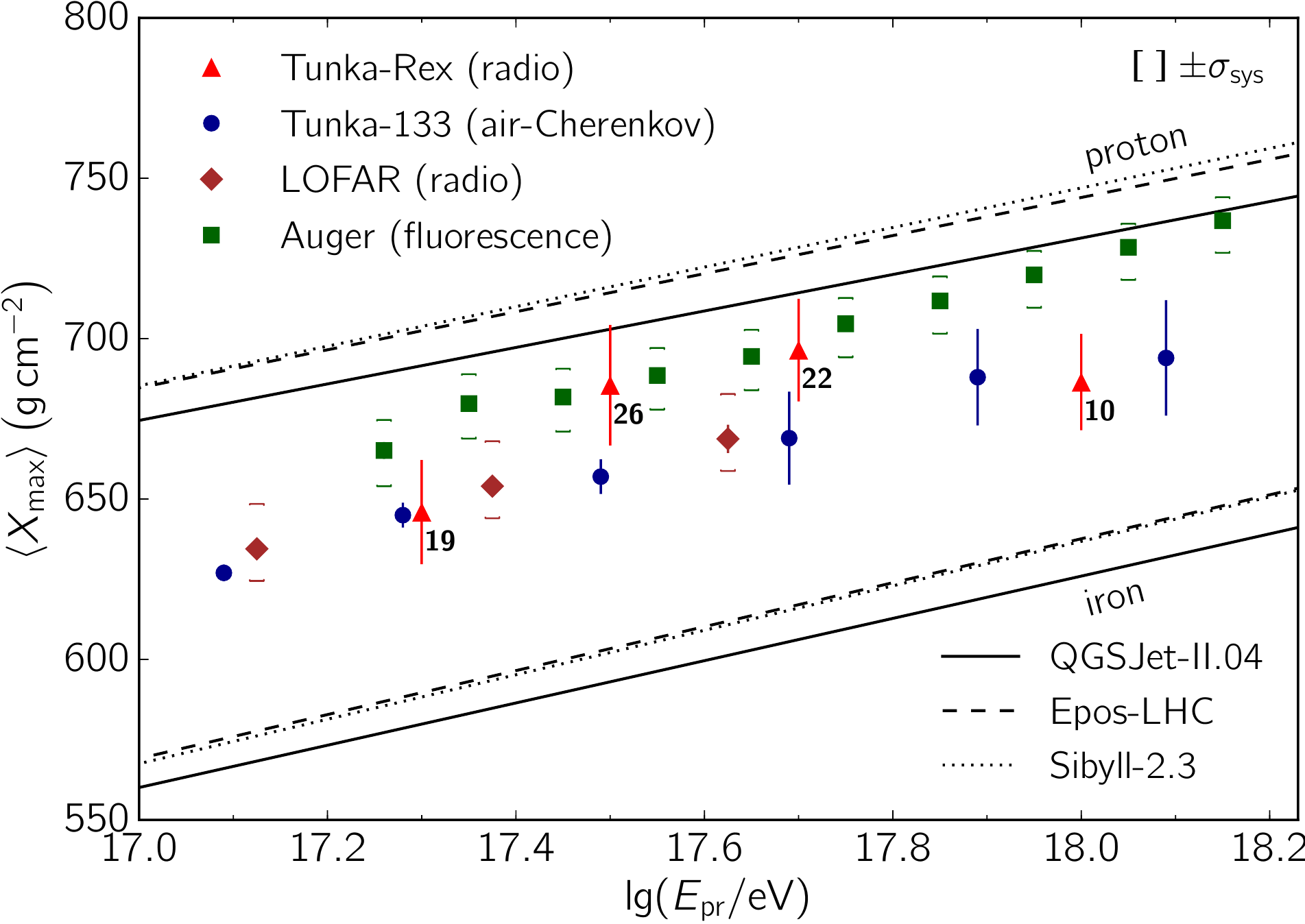}~~~
\includegraphics[height=0.35\textwidth]{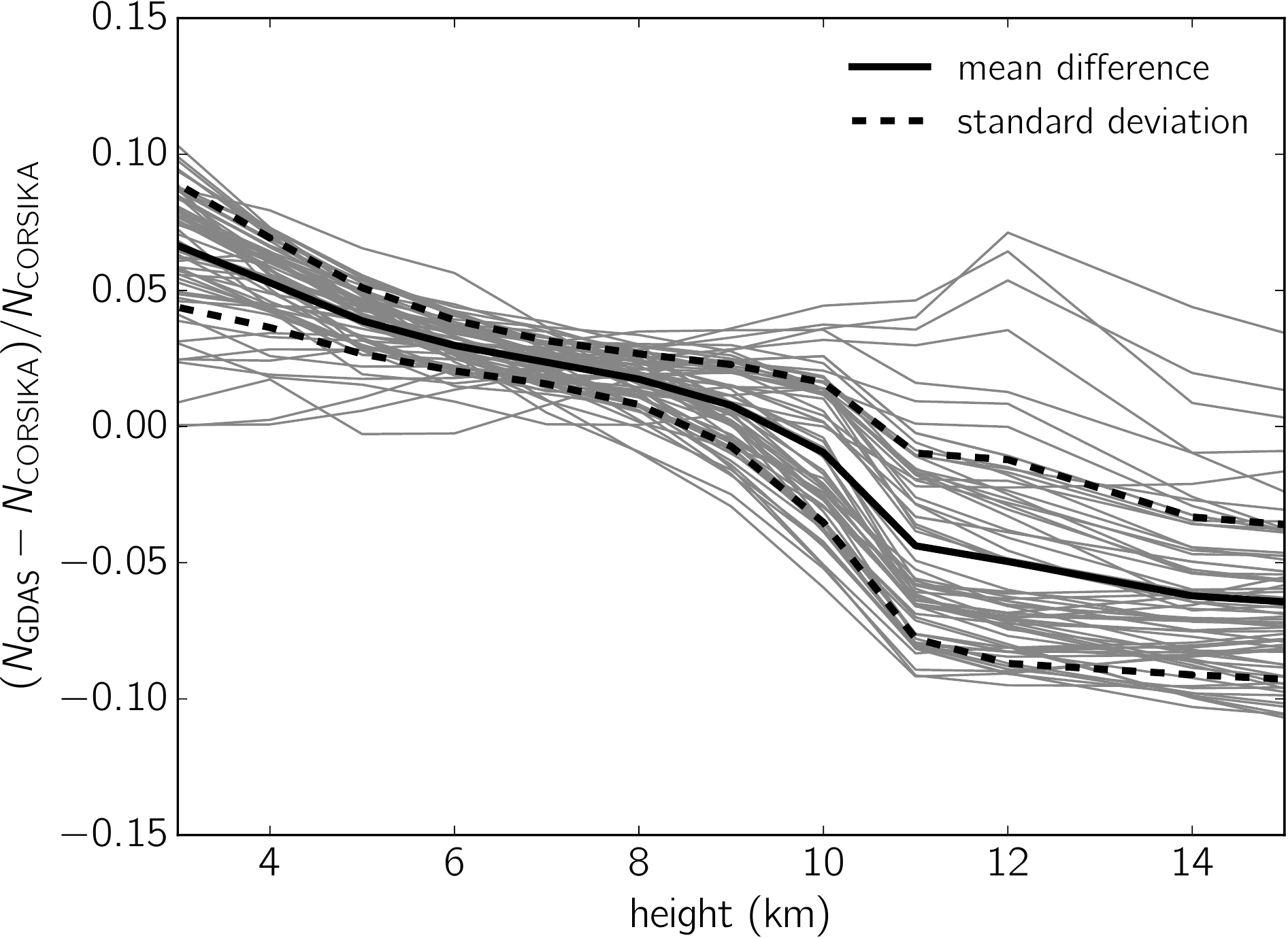}
\caption{\textit{Left:} mean shower maximum as function of primary energy reconstructed by different experiments measuring electromagnetic component of air-showers. 
\textit{Right:} difference between Tunka refractivity predicted with CORSIKA standard atmosphere and one calculated from Global Data Assimilation System (GDAS) profiles.}
\figlab{fig:xmax}
\end{figure}

Recently we applied template fit method to the Tunka-Rex reconstruction~\cite{Bezyazeekov:2018yjw}, which improved the reconstruction of the primary energy and depth of the shower maximum.
Contrary to the standard reconstruction by Tunka-Rex, which uses only the pulse maxima~\cite{Bezyazeekov:2015ica}, the new method additionally makes use of the pulse shape.
Based on the standard procedure the reconstructed events are reproduced with CoREAS~\cite{Huege:2013vt} simulations for different primary particles.
Then the pulse shapes of simulated radio pulses are fitted to the measured ones.
The combination of standard pre-reconstruction and template fitting can be summarized as follows:
\begin{enumerate}
\item Pre-reconstruction using the standard Tunka-Rex analysis pipeline with the improvements described above.
This reconstruction provides the shower axis and core position from Tunka-133 and the energy reconstructed by Tunka-Rex.
\item Creating a library with CoREAS simulations for each event obtained in the previous step with the goal to cover all possible depths of shower maxima possible for the particular event.
The reconstructed energy and geometry were used as input for the simulations with different primaries.
We use CORSIKA v75600~\cite{HeckKnappCapdevielle1998} with QGSJet-II.04~\cite{Ostapchenko:2010vb}.
\item Chi-square fit of the simulated envelopes against reconstructed ones.
The shower maximum and primary energy are reconstructed from the fits.
\end{enumerate}

Using the efficiency model developed for Tunka-Rex~\cite{Fedorov:2017xih}, we applied efficiency cuts on the reconstructed events and reconstructed mean shower maximum as a function of a primary energy (\figref{fig:xmax}, left).
To estimate the uncertainty introduced by the atmospheric model in CORSIKA, we compared model values with GDAS interpolation and found that for the subset selected for this work, uncertainty introduced by the atmosphere is $<5$~g/cm\textsuperscript{2} (\figref{fig:xmax}, right).

\begin{figure}[t]
\centering
\includegraphics[width=0.96\textwidth]{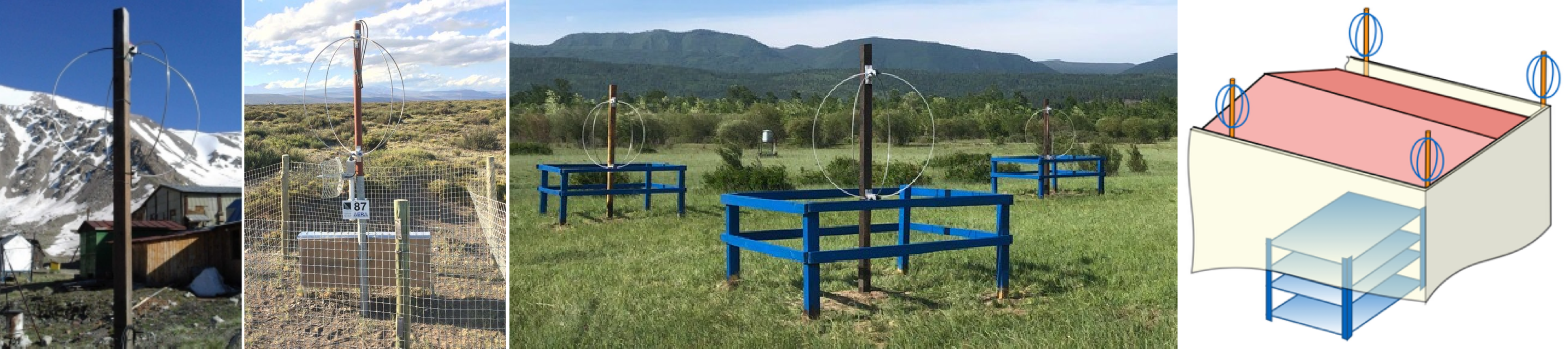}
\caption{Tunka-Rex SALLA installed at the different experiments. From left to right: Tien-Shan High Mountain Cosmic Station (Kazakhstan), Pierre Auger Observatory (Argentina)~\cite{Hoerandel_ARENA2018}, ISU educational cluster (Russia), future TRASGO detector (Spain)~\cite{Belver:2012zz}.}
\figlab{fig:tunkarex_collab}
\end{figure}

\section{Conclusion}
Tunka Radio Extension is a successful cosmic ray experiment operating since 2012.
The main results achieved by Tunka-Rex can be summarized as follows:
\begin{itemize}
\item Development robust methods of air-shower reconstruction with sparse radio arrays when each event contains few antenna stations with signal~\cite{Kostunin:2015taa, Bezyazeekov:2018yjw}.
\item Semi-blind cross-check of radio and air-Cherenkov techniques~\cite{Bezyazeekov:2015ica}.
\item Cross-calibration between Tunka-133 and KASCADE-Grande experiments using radio extensions~\cite{Apel:2016gws}, for which it was shown that, that the systematic shift in energy scales of these experiments is less than the uncertainty of energy reconstruction.
\item Development model describing aperture and exposure~\cite{Fedorov:2017xih} and application of this model for the reconstruction the mean shower maximum as function of primary energy~\cite{Bezyazeekov:2018yjw}.
\end{itemize}
Besides these main results, Tunka-Rex performed ultimate test of cost-effective and robust SALLA instrument, which was specially designed for cosmic-ray measurements.
Particularly, it was shown, that these antennas are sensitive to very inclined air-showers~\cite{Marshalkina_ARENA2018}.
In addition, Tunka-Rex collaboration shares this techniques with the partners, one can see the different setups exploiting SALLA in \figref{fig:tunkarex_collab}.

Let us mention the future plans and next milestones of the Tunka-Rex experiment.
\begin{itemize}
\item \textit{Energy spectrum with radio}. Although all-particle spectrum in energy range of 100-1000~PeV is measured with high statistics, there are still number of questions considering absolute flux of cosmic rays.
For example, combination of different radio experiments can help in the study of systematics as well as probe the North-South difference in absolute fluxes.
\item \textit{Mass composition combining radio and muons by Tunka-Grande} can improve sensitivity to the primary mass in Galaxy versus extra-Galaxy transition region in cosmic-ray spectrum.
\item \textit{Open data and software}. We plan to publish Tunka-Rex data and software in the frame of Russian-German Astroparticle Data Life Cycle initiative (visit \texttt{astroparticle.online} for details)
\end{itemize}

\section*{Acknowledgements} 
This work is supported by the Deutsche Forschungsgemeinschaft (DFG Grant No. SCHR 1480/1-1), 
the Russian Federation Ministry of Education and Science (Tunka shared core facilities, unique identifier RFMEFI59317X0005, 3.9678.2017/8.9, 3.904.2017/4.6, 3.6787.2017/7.8, 3.6790.2017/7.8), 
the Russian Foundation for Basic Research (Grants No. 16-02-00738, No. 17-02-00905, No. 18-32-00460). 
We used calculations performed on the HPC-cluster “Academician V.M. Matrosov” and on the computational resource ForHLR II funded by the Ministry of Science, Research and the Arts Baden-W\"urttemberg and DFG.

\bibliography{references}

\end{document}